\documentstyle[preprint,aps]{revtex}
\newcommand{\be}{\begin{equation}}
\newcommand{\ee}{\end{equation}}
\newcommand{\bea}{\begin{eqnarray}}
\newcommand{\eea}{\end{eqnarray}}
\newcommand{\ba}{\begin{array}}
\newcommand{\ea}{\end{array}}
\newcommand{\bc}{\begin{center}}
\newcommand{\ec}{\end{center}}
\newcommand{\btab}{\begin{tabular}}
\newcommand{\etab}{\end{tabular}}
\newcommand{\bt}{\begin{table}}
\newcommand{\et}{\end{table}}
\newcommand{\bitem}{\begin{itemize}}
\newcommand{\eitem}{\end{itemize}}
\newcommand{\dvec}[1]{\stackrel{\leftrightarrow}{#1}}

\begin{document}

\title{Hyperon-rich Matter in Neutron Stars}

\author{
J\"urgen Schaffner and Igor N. Mishustin\thanks{permanent address:
The Kurchatov Institute Russian Research Center, Moscow 123182, Russia}}
\address{
The Niels Bohr Institute\\
Blegdamsvej 17\\
DK-2100 Copenhagen}

\maketitle

\begin{abstract}
We study the equation of state of hyperon-rich matter for neutron stars
using an extended relativistic mean-field model.
We take special care
of the recently proposed non-linear behaviour of the vector field which
gives a much better description of
Dirac-Br\"uckner calculations.
The hyperon-hyperon interaction is also implemented
by introducing additional meson exchanges.
These new terms avoid the
instability found at high densitites in previous works while
keeping the excellent description for finite nuclear systems.
We also demonstrate within the mean-field approach
that the presence of hyperons
inside neutron stars on one hand and the hyperon-hyperon interactions
on the other hand make the onset of kaon condensation less favourable.
\end{abstract}

\section{Introduction}

Nuclear matter at high densities exhibits a new degree of freedom:
strangeness, hyperons and possibly kaons, occur at a moderate density of about
2-3 times normal nuclear density in neutron star matter \cite{Glen87}.
These new particles influence the properties of the equation of state
of matter and the global properties of neutron stars.
Usually the relativistic mean field (RMF) model is adopted for describing
the equation of state with hyperons.
But the standard approach suffers of several shortcomings:
First, the parametrizations adjusted to reproduce the
properties of nuclei fail at
high densities due to some instabilities of the scalar selfinteraction
(a possible solution to this problem has been suggested by Reinhard
\cite{Rei88}).
Second the equation of state is much stiffer compared to nonrelativistic or
Dirac-Br\"uckner calculations and the effective mass is much smaller
($m^*/m \approx 0.5-0.6$) in the ground state.
A possible way out is to introduce
a quartic selfinteraction term for the vector field \cite{Bod91}.
This model gives a good description of the properties of nuclei \cite{Toki94}
and is in reasonable agreement with Dirac-Br\"uckner calculations
\cite{Gmu91}.
In addition the instability found in the standard approach disappears.
The equation of state for neutron stars (without hyperons)
is considerably softened \cite{Toki94}.
Third the hyperon-hyperon interaction becomes important for hyperon-rich
matter present in the dense interior of neutron stars (note that nearly equal
amounts of hyperons and nucleons are predicted \cite{Glen87}).
The standard RMF model is not suited to reproduce the strongly attractive
hyperon-hyperon interaction seen in double $\Lambda$-hypernuclei.
An improved Lagrangian incorporating
an additional pair of (hidden) strange meson fields remedies the situation
\cite{Sch93,Sch94}.
These additional interaction terms have never been applied
for the equation of state of neutron star matter before.

In this paper
we study the equation of state of neutron star matter for representative
parameter sets which cover more or less all presently availabe fits
of the relativistic mean field model to the properties of finite nuclei.
The model is extended to include hyperons in a controllable way
by fitting the parameters to hypernuclear data.
In addition the hyperon-hyperon interaction is introduced through
(hidden) strange meson exchanges.

When describing the properties of dense matter it is also necessary to study
the possibility of pseudoscalar meson condensation. In particular, much
attention has been paid in recent years to the kaon condensation in neutron
stars \cite{Kap86}.
Most recent calculations based on chiral perturbation theory
\cite{Brown92,Brown94} show that kaon condensation may set in
at densities only 3--4 times larger than normal nuclear density.
Nevertheless, these calculations do not take into account
the presence of hyperons which will already occupy a large fraction of
matter when the kaons possibly start to condense. On the other side,
the calculations including hyperons have not taken into account the possible
kaon condensed phase.
Only recently, some work has been started to incorporate the hyperon and
kaon degrees of freedom at the same time \cite{Muto93,Ell95}.
In ref.~\cite{Muto93} the chiral Lagrangian by Kaplan and
Nelson \cite{Kap86} is used for the baryons and kaons.
The hyperons have not been included
explicitly as constituents of the ground state, they are only considered
through particle-hole excitations induced by the p-wave kaon-hyperon
interaction.
In this approach kaon condensation is predicted around $3\rho_0$.
The other work \cite{Ell95} uses the standard
relativistic mean field (RMF) approach for the baryon sector and
the kaon-baryon interactions
from the Kaplan-Nelson Lagrangian.
It is shown that the critical density
for kaon condensation is shifted to higher densities
when hyperons are included ($\rho_c > 4 \rho_0$).
It has been observed that the nucleon effective
mass is shifted towards negative values at high densities
due to the kaon condensate.
The authors conclude that
the RMF model breaks down at these densities and that one has to go beyond
the mean-field level.
Thus, two approaches using the same kaon-baryon interactions come
to different conclusions.
In view of the present interest
we also include the kaons in our RMF model by using
kaon-baryon interactions motivated by meson exchange models.
We demonstrate that the additional (hidden) strange meson
fields mentioned above
make kaon condensation less favourable even at very high densities.

\section{The extended RMF model}

We start from the standard Lagrangian
\bea
{\cal L} &=&
\sum_{\rm B} \overline{\Psi}_B(i \gamma^\mu\partial_\mu - m_B)\Psi_B
+ \frac{1}{2}\partial^\mu \sigma \partial_\mu \sigma
-  U(\sigma) \cr
&&
-\frac{1}{4}G^{\mu\nu}G_{\mu\nu}
+ \frac{1}{2}m_\omega^2 V^\mu V_\mu
-\frac{1}{4}\vec{B}^{\mu\nu}\vec{B}_{\mu\nu}
+ \frac{1}{2}m_\rho^2 \vec{R}^\mu \vec{R}_\mu \cr
&&
- \sum_{\rm B} g_{\sigma B}\overline{\Psi}_B\Psi_B\sigma
- \sum_{\rm B} g_{\omega B}\overline{\Psi}_B\gamma^\mu\Psi_B V_\mu
- \sum_{\rm B} g_{\rho B}\overline{\Psi}_B\gamma^\mu
                        \vec{\tau}_B\Psi_B\vec{R}_\mu
\eea
where the sum runs over all baryons of the baryon octet
(p,n,$\Lambda$,$\Sigma^+$,$\Sigma^0$,$\Sigma^-$,$\Xi^0$,$\Xi^-$).
The term $U(\sigma)$ stands for the scalar selfinteraction
\be
U(\sigma) = \frac{1}{2}m_\sigma^2 \sigma^2
+ \frac{b}{3}\sigma^3 + \frac{c}{4}\sigma^4
\label{eq:bognl}
\ee
introduced by Boguta \cite{Bog77} to get a correct
compressibility of normal nuclear matter.
The parameters of this Lagrangian have been fitted to the
properties of finite nuclei \cite{Rufa88,Shar93}.
It turned out that the best fits are obtained for the parameter sets with
$c<0$. In this case the functional form (\ref{eq:bognl}) of the scalar field
potential leads to an instability at high densities. Its traces are already
seen in the nucleus $^{12}$C \cite{Rei88}.
Hence, another stabilized functional form
has been given by Reinhard \cite{Rei88} which eliminates the instability
while keeping the good description of nuclei, especially the spin-orbit
splitting. Alternatively,
Bodmer proposed an additional selfinteraction
term for the vector field \cite{Bod91} of the form
\be
{\cal L}_{V^4} = \frac{1}{4} d (V_\mu V^\mu)^2
\quad .
\ee
This modification leads to a nice agreement with Dirac-Br\"uckner
calculations at high densities \cite{Gmu91}.
The reason is that the vector field is then proportional to $\rho^{1/3}$
in contrast to the linear dependence in the standard model.
The fits to the properties of nuclei are quite
succesful and the instability due to the scalar
selfinteraction vanishes \cite{Toki94}.

The implementation of hyperons is straightforward.
The corresponding
new coupling constants have been fitted to hypernuclear properties
\cite{Bro77}.
It turns out that the two coupling constants of the $\Lambda$
($g_{\sigma\Lambda}$ and $g_{\omega\Lambda}$) are strongly correlated
because they are fixed by the depth of the $\Lambda$-potential
\be
U_\Lambda^{(N)} =
g_{\sigma\Lambda}\sigma^{\rm eq.} + g_{\omega\Lambda} V_0^{\rm eq.}
\ee
in saturated nuclear matter \cite{Glen91,Sch92}. Here
$U_i^{(j)}$ denotes the potential depth of a baryon species $i$ in
matter of baryon species $j$.
Hence one can use
for example SU(6)-symmetry for the vector coupling constants
\be
\frac{1}{3}g_{\omega N} = \frac{1}{2} g_{\omega\Lambda}
= \frac{1}{2} g_{\omega\Sigma} =  g_{\omega\Xi}
\; , \qquad
g_{\rho N} = \frac{1}{2} g_{\rho\Sigma} =  g_{\rho\Xi} \; ,
\quad g_{\rho\Lambda} = 0
\ee
and fix the scalar coupling constants to the potential depth of the
corresponding hyperon.
Following \cite{Sch93,Sch94} we choose
\be
U_\Lambda^{(N)} = U_\Sigma^{(N)} = -30 \mbox{ MeV} \quad , \qquad
U_\Xi^{(N)} = -28 \mbox{ MeV}
\label{eq:potdep1}
\ee
in accordance with the available hypernuclear data.
This constitutes our model~1.

Nevertheless, these models are not able
to reproduce the observed strongly attractive $\Lambda\Lambda$ interaction
irrespectively of the chosen vector coupling constant.
The situation can be remedied
by introducing two additional meson fields, the scalar meson $f_0(975)$
(denoted as $\sigma^*$ in the following) and the vector meson
$\phi(1020)$ with the masses given in parenthesis \cite{Sch93,Sch94}.
The corresponding Lagrangian reads
\bea
{\cal L}^{YY} & = &
   \frac{1}{2}\left(\partial_\nu\sigma^*\partial^\nu\sigma^*
     - m_{\sigma^*}^2{\sigma^*}^2\right)
 - \frac{1}{4} S_{\mu\nu}S^{\mu\nu} + \frac{1}{2} m_\phi^2\phi_\mu\phi^\mu
\cr \cr
& &
 - \sum_B g_{\sigma^* B}{\overline \Psi}_B\Psi_B\sigma^*
 - \sum_B g_{\phi B}{\overline\Psi}_B\gamma_\mu\Psi_B \phi^\mu
\quad .
\eea
The vector coupling constants to the $\phi$-field are given by SU(6)-symmetry
(see \cite{Sch94} for details)
\be
2g_{\phi\Lambda} = 2g_{\phi\Sigma} = g_{\phi\Xi} = - \frac{2\sqrt{2}}{3}
g_{\omega N} \; , \quad g_{\phi N} = 0 \quad .
\ee
The scalar coupling constants to the
$\sigma^*$-field are fixed by the condition
\be
U^{(\Xi)}_\Xi \approx U^{(\Xi)}_\Lambda \approx
2U^{(\Lambda)}_\Xi \approx 2U^{(\Lambda)}_\Lambda \approx -40 \mbox{ MeV}
\label{eq:potdep2}
\ee
which is motivated by the one-boson exchange model D of the Nijmegen group
and the measured strong $\Lambda\Lambda$ interaction \cite{Sch94}.
Note that the nucleons are not coupled to these new fields.
In the following we denote the extended model with hyperon-hyperon interactions
as model 2.

\section{Neutron star matter with hyperons}

The equation of state for neutron star matter is derived by standard methods
(see e.g.\ ref.~\cite{Glen87} for the standard RMF approach).
The equations of motion for the meson fields in uniform matter are given by
\bea
m_\sigma^2 \sigma + \frac{\partial}{\partial \sigma} U(\sigma ) &=&
\sum_{\rm B} g_{\sigma B} \rho^{(B)}_S \cr
m_{\sigma^*}^2 \sigma^* &=&
\sum_{\rm B} g_{\sigma^* B} \rho^{(B)}_S \cr
m_\omega^2 V_0 + d V_0^3 &=&
\sum_{\rm B} g_{\omega B} \rho^{(B)}_V \cr
m_\rho^2 R_{0,0} &=&
\sum_{\rm B} g_{\rho B} \tau_3^B \rho^{(B)}_V \cr
m_\phi^2 \phi_0 &=&
\sum_{\rm B} g_{\phi B} \rho^{(B)}_V
\quad .
\eea
where $\rho_S$ and $\rho_V$ denote
the scalar and vector densities, respectively.
The equations can be solved for a given total baryon density $\rho_B$ and
charge density $\rho_c$
including the contributions from the free electrons and muons
\bea
\rho_B &=& \sum_{\rm B} \rho^{(B)}_V \cr
\rho_c &=& \sum_{\rm B} q_B \rho^{(B)}_V + \sum_{l=e,\mu} q_l \rho_l = 0
\eea
where $q_i$ stands for the electric charge of a species $i$.
In $\beta$-equilibrium the chemical potentials of the particles
are related to each other by
\be
\mu_i = B_i \cdot \mu_B + q_i \cdot \mu_e
\ee
where $B_i$ is the baryon number of a species $i$.
This means that all reactions which
conserve charge and baryon number are allowed, as e.g.\
\be
{\rm n} + {\rm n} \to \Lambda + {\rm n} \quad , \qquad
\Lambda + \Lambda \to \Xi^- + {\rm p} \quad, \qquad \dots
\ee
Since we consider neutron stars in a long time scale, the strangeness quantum
number is not constrained and the net strangeness is determined by the
condition
of $\beta$-equilibrium.
The above equations fixes already the fields and the composition of
neutron star matter.
The energy and pressure density
can be derived from the grand canonical potential
or the energy-momentum tensor in a standard way (see e.g.\ \cite{Glen87})
and the generalization to the additional fields is straightforward:
\bea
\epsilon &=& \frac{1}{2}m_\sigma^2 \sigma^2
+ \frac{b}{3}\sigma^3 + \frac{c}{4}\sigma^4
+ \frac{1}{2}m_{\sigma^*}^2 {\sigma^*}^2
+ \frac{1}{2}m_\omega^2 V_0^2 + \frac{3}{4} d V_0^4
\cr && {}
+ \frac{1}{2}m_\rho^2 R_{0,0}^2
+ \frac{1}{2}m_\phi^2 \phi_0^2
+ \sum_{i=B,l} \frac{\nu_i}{(2\pi^3)} \int_0^{k_F^i} d^3 k
\sqrt{k^2 + {m^*_i}^2} \\ \cr
P &=& - \frac{1}{2}m_\sigma^2 \sigma^2
- \frac{b}{3}\sigma^3 - \frac{c}{4}\sigma^4
- \frac{1}{2}m_{\sigma^*}^2 {\sigma^*}^2
+ \frac{1}{2}m_\omega^2 V_0^2 + \frac{1}{4} d V_0^4
\cr && {}
+ \frac{1}{2}m_\rho^2 R_{0,0}^2
+ \frac{1}{2}m_\phi^2 \phi_0^2
+ \sum_{i=B,l} \frac{\nu_i}{(2\pi^3)} \int_0^{k_F^i} d^3 k
\frac{k^2}{\sqrt{k^2 + {m^*_i}^2}}
\eea
where
\be
m^*_B = m_B + g_{\sigma B}\sigma + g_{\sigma^* B} \sigma^*
\label{eq:efmas}
\ee
is the effective mass of the baryon $B$ (for leptons the on-shell mass
is taken).

In the following we study
the composition of neutron star matter for the various parameter
sets obtained from fits to finite nuclei.
First we take the sets NL-Z (which is  set NL-1 with a more consistent
center-of-mass correction \cite{Rufa88}), NL-SH \cite{Shar93}
for the Boguta-model and the sets PL-40 and PL-Z \cite{Rei88}
with the stabilized scalar functional. The corresponding coupling constants
are listed in table~1.
The first parameter set shows a well known instability at high densities
\cite{Rei88} which appears in neutron star matter around
$\rho \approx 0.5$~fm$^{-3}$.
At this point the effective mass of the nucleon gets negative
due to the presence of the hyperons.
For higher density no solution can be found.
Negative effective nucleon masses appear also for set NL-SH at a
density of $\rho \approx 0.8$~fm$^{-3}$
for model 1 and at 1.0~fm$^{-3}$ for model 2.
The same holds for the case with the stabilized functional forms, sets PL-Z and
PL-40 \cite{Rei88}, in model 1 and 2 around a similar density region.
Note, that this happens even without the presence of a kaon condensate
as found in \cite{Ell95}.
One might wonder about the fact, that the
early occurence of negative masses found here has not been seen in
\cite{Ell95}. The explanation is simple: The standard parameter sets from
fits to finite nuclei always favour a stiff equation of state to
get a correct spin-orbit splitting which inhibits this instability as
the fields are growing fast with the density.
Fits to nuclear matter only can lead to a rather soft equation of state
which avoids the instability due to a moderate rise of the fields.
Nevertheless, the latter parametrizations are not
able to describe the strong spin-orbit splitting seen in finite nuclei
(for a detailed discussion see \cite{Rufa88}) and therefore can not be
considered as realistic. In the following
we extended our calculation to densities beyond the instability,
by taking always the absolut value $|m_N^*|$ of the effective mass
(it is clear, that a much more refined procedure is necessary to treat
the problem of negative effective masses, i.e.\
going beyond the mean-field appoximation).

The behaviour for lower densities is quite similar in all these models.
Fig.~1 shows as an example the composition of neutron star
matter for the set PL-Z and model 2.
The proton fraction raises
rapidly and reaches maximum values over 20\% around $2-3\rho_0$.
At this density, the hyperons, first $\Lambda$'s and $\Sigma^-$'s then
$\Xi^-$'s, appear abundantly and the lepton fraction is considerably reduced.
When the other hyperons are present at densities of $3-4\rho_0$
the number of $\Lambda$'s even exceeds the number of neutrons, so that the
dense interior resembles more a hyperon star than a neutron star
\cite{Glen87}.
Also the population of leptons gets negligibly low, as the electrochemical
potential drops instead of growing.
We have also checked for the occurence of
$\Omega^-$ which do not contribute in any of our calulations.

The situation for higher densities differs for the various parameter sets
mainly due to the instability mentioned above.
Taking the absolute value of the effective nucleon mass still
no solution can be found for the sets NL-Z and NL-SH above the critical
density for model 1.
For the sets PL-Z and PL-40 we found a solution and
the neutron star matter becomes pure hyperon matter
dominated by $\Lambda$'s and no nucleons appear above the critical density.
Also the electrochemical potential changes sign
and positrons and antimuons appear instead of electrons and muons.
When turning on the hyperon-hyperon interactions (model 2)
the hyperons feel at these high densities effectively an additional repulsion.
Hence, nucleons are now present but still considerably
less abundant than hyperons.
Now a solution for set NL-SH can be also found.
Nevertheless, these results should be taken with some care as
we do not treat the problem of negative effective masses consistently.

Now we turn our discussion to the case of models with vector selfinteractions.
Two sets, named TM1 and TM2, exist so far in the literature \cite{Toki94}
(see also table~1).
It has already been found that the behaviour
of the nuclear equation of state follows more nicely the trends of relativistic
Br\"uckner-Hartree-Fock calculations and that the maximum mass
of a neutron star is reduced \cite{Toki94}.
Fig.~2 shows now the composition of neutron star matter for the set TM1 with
hyperons and with hyperon-hyperon interactions (model 2).
Up to the maximum density considered, here $10\rho_0$, all effective masses
remain positive and no instability occurs. The behaviour at moderate densities
is quite close to that of the previous parametrizations: The proton fraction
has a plateau at $2-4\rho_0$ and exceeds 20\%{} which allows for the
direct URCA process and a rapid cooling of the neutron star \cite{Lat91}.
Hyperons, first $\Lambda$'s and $\Sigma^-$'s, appear at $2\rho_0$, then
$\Xi^-$'s are populated already at $3\rho_0$. The number of electrons
and muons has a maximum here. The muons vanish when all the hyperons
have been settled at $\rho\approx 0.85$~fm$^{-3}$ while the electrons remain
on the level of 2\%{}. The fractions of all baryons show a tendency towards
saturation, they reach asymptotically similar values of $8-15 \%$
which corresponds approximately
to spin- and isospin-saturated matter. In the ideal case all baryons fractions
would be the same due to the same spin-degeneracy factors.

Switching off the hyperon-hyperon interactions, i.e.\ going from model 2 to
model 1, we see again that negative
effective masses occur at $\rho = 1.3$~fm$^{-3}$ for set TM1 and at
$\rho = 0.95$~fm$^{-3}$ for set TM2 due to the missing additional
repulsive force for the hyperons. Nevertheless, the main population pattern is
not changed considerably below $\rho = 0.6$~fm$^{-3}$ when going from model 2
to model 1, but the number of hyperons, especially $\Lambda$'s, are
reduced at higher densities. Also the leptons vanishes for set TM1,
but for the set TM2 positrons appear again,
as the electrochemical potential changes sign above the occurrence of
the instability.
These results demonstrate the importance of the additional exchanges of
$\sigma^*$ and $\phi$ mesons.
An additional repulsion is needed at higher densities to stabilize neutron star
matter.

The effective masses defined in (\ref{eq:efmas})
are plotted in fig.~3 for the various baryons for set TM1 and model 2.
The effective mass of the
nucleons reaches rapidly very small values and then saturates at higher
densities never reaching negative values.
For hyperons the behaviour is quite similar but less pronounced
as their coupling constants to the $\sigma$-field are smaller than for
nucleons.
Nevertheless, the combined effect of the $\sigma$ and $\sigma^*$ fields
results in a rather constant gap between the effective masses over the
whole density range shown.
Note that the nucleons do not couple to the (hidden)
strange scalar field $\sigma^*$.
The overall small masses of the baryons at very high densities indicate
that the neutron star matter approaches isospin-saturated matter with
equal abundances, as can be seen in fig.~2.

The pure scalar and vector field potentials are shown in fig.~4
(they have to be multiplied with the corresponding coupling constants given
in table~1 to get the potential for a baryon species).
In addition, we have also plotted the electrochemical potential.
It reaches a maximum value of about 200 MeV around $\rho\approx 2-3\rho_0$
and gets then consideraly lower at high
densities instead of growing steadily. This behaviour is due to the onset
of negatively charged hyperons ($\Sigma^-$ and $\Xi^-$).
The vector potential exhibits initially a linear rise which then slows down due
to the vector selfinteraction and the onset of hyperons at
$\rho\approx 0.3$~fm$^{-3}$.
At the same density the (hidden) strange fields are developed and reach quite
high values for very dense matter.
The $\phi$-field shows a rather linear behaviour with density while
the scalar fields seems to saturate at high densities.
The isovector potential coming from
$\rho$-exchange is quite small over the whole density range and never exceeds
$-60$ MeV for the nucleons.
The other potentials are one order of magnitude higher and have opposite signs.
That also indicates a large cancelation between scalar and vector terms.
The fields get so strong, that, for example, a 10\% change of one coupling
constant can alter the potential up to 100 MeV!
Therefore, it is very important to fine tune the scalar coupling constants
according to the potential depth
(see eqs.~(\ref{eq:potdep1}) and (\ref{eq:potdep2})).
This has to be kept in mind for the forthcoming discussion
of the properties of kaons in dense neutron star matter.

The equation of state, the pressure density versus the energy density is given
in fig.~5 for all parameter sets using model 2.
For $\epsilon < 800$~MeV$\cdot$fm$^{-3}$ there exist mainly two bunches of
curves.
The parameter sets without vector selfinteraction are located in the upper
branch.
The sets TM1 and TM2 with vector selfinteraction are
softer, i.e.\ they have a lower pressure for a given
energy density and are therefore constituting the lower branch of curves.
For higher energy densities, set TM2 shows to be stiffer
than set TM1.
For the upper branch one recognizes kinks in the curves which are due to
the instability discussed above (i.e.\ negative effective masses).
For model 1 (not shown), one sees a very
pronounced jump in the curves due to the
occurence of this instability.
The equation of state gets then considerably softened reaching very low
pressures.
But for model 2 the jump is much less pronounced and even vanishes for the
sets with vector selfinteraction.
Another difference between model 1 and model 2 is the slight
softening of the equation of state for medium densities which is due
to the attractive hyperon-hyperon interaction mediated by the $\sigma^*$-meson
exchange. For very high densities model 2 gets stiffer than model 1, because
the contribution from the repulsive $\phi$-field overwhelms now the
one coming from the attractive
$\sigma^*$-field which is saturating at very high densities.
In addition the stiffest possible equation of state $\epsilon = p$ is also
drawn. The causality condition $\partial p/ \partial\epsilon\leq 1$ is
fulfilled by all sets, so that the speed of sound remains lower than the
velocity of light. The microscopic stability condition
$\partial p/ \partial\epsilon\geq 0$ is also satisfied except for the
instability region.

Finally, we want to study possible extensions of the RMF approach
including other mesons. It is well known that pseudoscalar and pseudovector
mesons do not contribute in the mean-field approximation. The same holds
for all the kaon resonances as they couple off-diagonally. Tensor mesons
also vanishes in infinite matter. Hence, the only remaining meson which can
be added to the present Lagrangian
is the scalar isovector meson $a_0(980)$ (the $\delta$-meson).
To our knowledge this meson has been not considered so far for the equation
of state of neutron stars. Fits to the properties of nuclei seems to indicate
that its contribution are negligible for the discussion about binding
energy, radius and surface thickness of stable nuclei \cite{Rufa88}.
Nevertheless, it might be important for very asymmetric systems.
For the discussion in the next section it is quite useful to study
the influence of this meson for the properties of neutron star matter.
We implement the $\delta$-meson in the standard way and choose
the coupling constant from the Bonn-model ($g_{\delta N}=5.95$
\cite{Mach87}).
The other coupling constants are
scaled according to the isospin of the corresponding baryon
($2g_{\delta N} = g_{\delta \Sigma} = 2g_{\delta \Xi}\,$,
$g_{\delta \Lambda}=0$). As seen in fig.~4 the contribution from the
vector isovector meson $\rho$ is quite negligible compared to the other
fields. Therefore, we expect only minor changes when introducing the
scalar isovector meson $\delta$. Indeed, the equation of state with the
$\delta$-meson gets only a little bit stiffer for lower density regions.
For higher densities symmetric matter is reached also and
the equation of state approaches the one without the $\delta$-meson
because the contribution of this meson
simply vanishes in symmetric matter.
The baryon composition shows some small changes for densities lower than
$\rho<3\rho_0$. As the $\delta$-field is repulsive for the protons, but
attractive for the $\Sigma^-$'s, the former ones are
a little bit suppressed while the
latter ones appear slightly earlier. These effects reduce the number of
electrons and the maximum chemical potential is now about $150$ MeV instead
of $200$ MeV. The meson fields do not change when the $\delta$-field is
introduced.
The most pronounced effect, despite of the lower electrochemical potential,
is the change in the effective masses of the baryons
which are plotted in fig.~6.
In contrast to fig.~3 every baryon has now a different effective mass.
The biggest effect is seen for the $\Sigma^-$ due to its large coupling
constant to the $\delta$-field. The effective mass of the $\Sigma^-$ can even
be lower than that of the $\Lambda$. The neutron has a significantly lower
effective mass than that of the proton. For high densities it can be
a relative difference of a factor of two! The absolute difference can be
as high as 120 MeV.
Nevertheless, the potential coming from the $\delta$-field never exceeds
70 MeV for the nucleons and is therefore negligible compared to the rather
high potential terms coming from the isoscalar fields (see fig.~4).

\section{Kaon interaction in the RMF approach}

The kaon-nucleon interactions has
been studied recently within different approaches
including the Bonn model \cite{Bonn90},
the Nambu-Jona-Lasinio model \cite{Lutz92}
and chiral perturbation theory \cite{Lee94}.
Antikaons in dense nuclear matter are of special interest because
they can form a condensate \cite{Kap86} leading to a considerable softening of
the equation of state and a reduction of the maximum mass of neutron
stars \cite{Brown92}. But this issue is still controversial. For instance
rather strong nonlinear behaviour has been found by extrapolating
KN-interactions derived from KN-scattering to dense nuclear matter
in the NJL-model \cite{Lutz92} where
no hints for kaon condensation is seen.
Applying the Bonn interaction scheme within the simple RMF approximation
also leads a strong nonlinear density dependance disfavouring antikaon
condensation \cite{Sch94b}.
Still, chiral perturbation theory gives a rather robust prediction for the
onset of antikaon condensation at $\rho_c \approx 3-4\rho_0$
\cite{Brown94}.

We highly appreciate many efforts devoted to the consistent approach to kaon
condensation based on the chiral perturbation theory
\cite{Brown94,Lee94,Lee95}.
But in our opinion this approach has principal difficulties which may not be
overcome in the near future. The main problem is that simple chiral models do
not describe correctly the saturation properties of nuclear matter. By this
reason the parameters used for the calculation of the kaon energy (coupling
constants, effective masses, etc.) are not consistent with the properties of
the
background nuclear matter. Another and maybe related problem is that only
linear
terms in the baryon density are under control in the chiral approach.
Therefore,
all nonlinear terms are dropped out. In particular, the scalar density $\rho_s$
is identified with the baryon (vector) density $\rho$ of the matter. This
approximation certainly fails at high densities (for instance at $\rho\approx
3\rho_0$ the difference can be about 80\% \cite{Rei88}).
Already using the scalar density, as it is dictated by Lorentz invariance,
shifts the kaon condensation treshold to densities above $5\rho_0$
\cite{Sch94b,Maru94}. In ref.~\cite{Sch94b} it was also shown that many-body
forces associated with the selfinteraction of the meson fields disfavour kaon
condensation even more. By these reasons we use a more phenomenological
approach
based on the one-boson exchange model in the present paper.

As it is well known, there exists an important difference between
K$^+$N and K$^-$N-interaction.
In contrast to K$^+$N-scattering the situation with K$^-$N-inter\-actions
is more complicated due to
the existence of the $\Lambda (1405)$-resonance just below treshold.
This makes the interpretation of the available K$^-$N-scattering
and K$^-$-atomic data less transparent.
Recently an improved fit of K$^-$-atomic data was carried out assuming
a nonlinear density dependence of the effective $t$-matrix \cite{Fried93}.
It has been shown that
the real part of the antikaon optical potential can be as attractive as
\be
U^{\bar K}_{\rm opt} \approx -200 \pm 20 \mbox{ MeV}
\ee
at normal nuclear matter density while being slightly repulsive at
very low densities. This is in accordance with the low density theorem
and K$^-$p-scattering and effected by the $\Lambda (1405)$-resonance.
Also another family of solutions have been
found with a moderate potential depth around $-50$~MeV.
Note that also the standard linear extrapolation gives only values of
about $-85$~MeV \cite{Fried93}. These latter two solutions
do not fulfill the low
density theorem, i.e.\ getting repulsive at low densities.
There exist some hints that the $\Lambda (1405)$ is a quasi-bound
state in the t-channel \cite{Sie88,Bonn90}.
It is not surprising then that the K$^-$N-scattering data can be explained by a
simple vector meson exchange \cite{Sie88} where
the effects of this resonance come out automatically.
This resonance seems to be less important in dense matter when the
kaon mass is shifted down below $m_{\Lambda(1405)}-m_N\approx 465$ MeV.
In ref.~\cite{Koch94} a separable potential was applied for the
K$^-$p-interaction for finite density.
Indeed, it was found that the mass of the $\Lambda(1405)$
is shifted upwards and exceeds the K$^-$p threshold already at densities of
about $\rho\approx 0.4\rho_0$.
In this case the use of mean-field potentials may be justified.
Most recently, the $\Lambda (1405)$ has been also included
in the chiral approach \cite{Lee95}.
No effect has been found for the onset of kaon condensation \cite{Lee95},
but note that the $\Lambda(1405)$ has been only implemented
as a new field contrary to the conclusion drawn in
\cite{Sie88,Bonn90}.
In a recent preprint the coupled channel analysis
of Siegel and Weise \cite{Sie88} has been applied for chiral perturbation
theory \cite{Kai95}. The coupled channel formalism automatically generates
the $\Lambda(1405)$ and succesfully describes the low energy K$^-$p-scattering
data.

In our previous work \cite{Sch94b} we have conjectured an important role which
hyperons should play in determing the threshold of kaon condensation. Here we
present the calculations which confirm these qualitative estimates.
Recently, the effects of the presence of hyperons has been studied also in
ref.~\cite{Ell95}.
The authors use the RMF model with scalar selfinteractions
for the baryon sector and the chiral approach of Kaplan and Nelson \cite{Kap86}
for the kaon-baryon interaction.
It has been shown that the presence of hyperons, particular the $\Sigma^-$,
shifts the threshold for K$^-$ condensation by several units of $\rho_0$. Note,
however, that Ellis et al.\ do not take into account the terms which ensure
that
the $\Sigma$-term changes sign at low
densities, as it should \cite{Brown94}.
These additional terms will shift the onset of
kaon condensation to even higher densities.

In the following
we adopt the meson-exchange picture for the KN-interaction simply
because we use it also for parametrizing the baryon interactions.
By analogy to the Bonn model \cite{Bonn90} the simplest kaon-meson Lagrangian
can be written as
\bea
{\cal L}_{K} &=& \partial_\mu \bar K \partial^\mu K - m_K^2 \bar K K
- g_{\sigma K} m_K \bar{K}K \sigma
- g_{\sigma^* K} m_K \bar{K}K \sigma^*
\cr &&
- g_{\omega K}\bar{K}i\dvec{\partial_\mu} K V^\mu
- g_{\rho K}\bar{K}\vec{\tau}_K i\dvec{\partial_\mu} K \vec{R}^\mu
- g_{\phi K}\bar{K}i\dvec{\partial_\mu} K \phi^\mu
\label{eq:knbonn}
\quad .
\eea
However, it is easy to see that this coupling scheme
destroys the gauge invariance of the vector interactions represented by the
equation
\be
\partial_\mu V^\mu = 0
\quad .
\ee
{}From the equation of motion for the vector field $V^\mu$ one has
\be
\partial_\mu F^{\mu\nu} + m_\omega^2 V^\nu =
\sum_{\rm B} g_{\omega B}\overline{\Psi}_B\gamma^\nu\Psi_B
+ g_{\omega K}(\bar{K}i\dvec{\partial_\nu} K)
\ee
and after applying a partial derivative on both sides
\be
m_\omega \partial_\nu V^\nu =
\sum_{\rm B} g_{\omega B}\partial_\nu (\overline{\Psi}_B\gamma^\nu\Psi_B)
+ g_{\omega K}\partial^\nu (\bar{K}i\dvec{\partial_\nu} K)
\ee
(the vector selfinteraction term is omitted here because
it is purely phenomenological and
the following arguments do not hold for it).
The first term on the right hand side vanishes due to the baryon number
conservation. The second term would also vanish for a free kaon, but
it does not in the medium. Indeed,
the conserved kaon current from the Lagrangian (\ref{eq:knbonn}) reads
\bea
j_\mu^K &=& i \left( \bar K \frac{\partial{\cal L}}{\partial^\mu \bar K}  -
\frac{\partial{\cal L}}{\partial^\mu K} K \right) \cr \cr
&=& \bar K i \partial_\mu K - (i \partial_\mu \bar K) K
+2\bar K K (g_{\omega K} V_\mu + g_{\rho K} \vec{\tau}\vec{R}_\mu + g_{\phi K}
\phi_\mu)
\eea
so that this term does not vanish if the expectation value of the vector fields
are nonzero. The situation can be corrected by introducing
the additional terms
\be
{\cal L}'_K = {\cal L}_K +
(g_{\omega K} V_\mu + g_{\rho K} \vec{\tau}\vec{R}_\mu + g_{\phi K}\phi_\mu)^2
\bar K K
\label{eq:knadd}
\quad .
\ee
The kaon-meson Lagrangian can now be written as
\be
{\cal L}'_K = D^*_\mu \bar K D^\mu K - m_K^2 \bar K K
- g_{\sigma K} m_K \bar{K}K \sigma
- g_{\sigma^* K} m_K \bar{K}K \sigma^*
\ee
with the covariant derivative
\be
D_\mu = \partial_\mu +
ig_{\omega K} V_\mu + ig_{\rho K} \vec{\tau}\vec{R}_\mu + ig_{\phi K}\phi_\mu
\quad .
\ee
The equation of motion for kaons in uniform matter then reads
$$
\left\{ \partial_\mu\partial^\mu + m_K^2
+ g_{\sigma K} \sigma m_K + g_{\sigma^* K} \sigma^* m_K
+ 2(g_{\omega K} V_0 + g_{\rho K} \tau_0 R_{0,0} + g_{\phi K}\phi_0)
 i\partial^\mu \right.
$$
\be
\left.
- (g_{\omega K} V_0 + g_{\rho K} \tau_0 R_{0,0} + g_{\phi K}\phi_0)^2
\right\} K = 0
\label{eq:KGkaon}
\quad.
\ee
The coupling constants to the vector mesons are chosen from the
SU(3)-relations
\be
2g_{\omega K} = 2g_{\rho K} = \sqrt{2} g_{\phi K} = g_{\pi\pi\rho} = 6.04
\quad .
\ee
Decomposing the kaon field into plane waves one ontains the following
dispersion
relation for kaons (upper sign) and antikaons (lower sign) in uniform matter
composed of nucleons and hyperons
\be
\omega_{K,\bar K} =
\sqrt{m_K^2 + m_K(g_{\sigma K} \sigma + g_{\sigma^* K}\sigma^*) + k^2 }
\;\pm\; \left( g_{\omega K} V_0 + g_{\phi K} \phi_0 + g_{\rho K} R_{0,0}
\right)
\label{eq:ekaon}
\quad .
\ee
Note that due to the additional terms (\ref{eq:knadd}) the
vector term appears linearly in the kaon energy.

The coupling constant $g_{\sigma K}$ could be taken from the Bonn model
\cite{Bonn90} but here we adopt another prescription. We want to fix it to
the value of the potential depth
of kaons in the nuclear medium to study the strongly attractive potential
seen in kaonic atoms \cite{Fried93}. The optical potential
in symmetric nuclear matter ($R_{0,0}=0$, $\phi_0=0$, no hyperons, $k=0$)
is in our approach given by
comparison with eq.~(\ref{eq:KGkaon})
\be
2\mu_{KN} U^{\bar K}_{\rm opt} = g_{\sigma K}\sigma m_K
   - 2 g_{\omega K} \omega_{\bar K} V_0 - (g_{\omega K} V_0)^2
\label{eq:knopt}
\ee
where $\mu_{KN}$ is the reduced mass of the K$^-$N-system.
One notices that in general the optical potential and the energy shift
(relative
to the free mass) of the (anti)kaon do not coincide.
In addition, the right hand side of eq.~(\ref{eq:knopt})
shows a nonlinear density dependance (see also \cite{Fried93}).
At normal nuclear density the in-medium energy shift of
the antikaon is about 20\% less than the corresponding optical potential!
In the following we study two possibilities: we adjust the scalar
coupling constant to $U^{\bar K}_{\rm opt}= -120$ MeV,
as an upper bound for the
standard fit and the lower family of solutions found in \cite{Fried93},
and to $U^{\bar K}_{\rm opt}= -200$~MeV for the second family of solutions.

For the $\bar K N$ case the low density theorem is not fulfilled
due to the presence of the $\Lambda (1405)$ resonance
(in principle one has to use a coupled-channel formalism to get the correct
scattering length which is beyond the scope of this paper).
We argue here that the repulsive regime at very low density is of no
consequence for the evaluation of K$^-$ nuclear interaction at high densities
as pointed out in \cite{Fried93,Koch94}.

The onset of (s-wave) kaon condensation is determined by the condition
\be
- \mu_e = \mu_{K^-} = \omega_{K^-} (k=0)
\quad .
\ee
When calculating $\mu_{K^-}$ we have also taken into account the
contribution from the (hidden) strange meson fields $\sigma^*$ and $\phi_0$
according to eq.~(\ref{eq:ekaon}).
The coupling constant to the $\sigma^*$ field has been taken from
$f_0$-decay \cite{Arm91} (note that this value is obtaint despite of the fact
that the mass of the $f_0$ is lower than the $K\bar K$ treshold which
constitutes another source of uncertainty).
While the $\sigma^*$-field is attractive, the $\phi$-field is repulsive
for antikaons.
The isovector field is also repulsive for the K$^-$
as neutrons (and also $\Sigma^-$) are more abundant than protons ($\Sigma^+$).
As seen in fig.~5 the electrochemical potential decreases for higher
densities. The relativistic effects, the reduced electrochemical
potential, and the presence of the $\phi$- and the $\rho$-meson fields
make the kaon condensation less favourable
for strangeness-rich matter.

In fig.~7 we plot $\omega_{K^-} - \mu_{K^-}$ as a function of baryon density
for model 2 and the various parameter sets considered. The upper curves
correspond to $U^{\bar K}_{\rm opt} = -120$~MeV whereas the lower curves to
$U^{\bar K}_{\rm opt} = -200$~MeV.
Below $2\rho_0$ the curves drop rather steep and linear.
After the appearance of the hyperons,
the slope of the curves changes dramatically
and they even turn upward again. Antikaon condensation is only possible if the
curves crosses zero. This does not happen at all for the upper curves
and only occurs for set PL-40 (but in the unstable regime
with negative effective masses!) and for set
TM1 among the lower curves. Note that the two sets TM1 and TM2, which do not
get negative effective masses and might be the only two reliable
calculations, give different predictions for the possible existence of a kaon
condensed phase. While the former one reaches zero at $\rho \approx 6\rho_0$,
the latter one is at all densities at least 60~MeV above the critical value.

For the case of $U^{\bar K}_{\rm opt} = -200$~MeV we have the problem
that the optical potential of the kaon turns out to be attractive
while experiment tells us that it is repulsive.
Also the low density theorem is not fulfilled in our model.

Hence, we discuss in the following the
KN-scattering lengths, try to fix our parameters
to the KN-scattering lengths and perform a G-parity transformation afterwards
to get the $\bar{K}$N-case.
The isospin averaged scattering length in the tree approximation
reads
\be
\bar{a}_0 = \frac{1}{4} a_0^{I=0} + \frac{3}{4} a_0^{I=1} =
\frac{m_K}{4\pi \left( 1+ m_K/m_N \right)} \left(
\frac{ g_{\sigma K}g_{\sigma N} }{ m_\sigma^2 }
-2\frac{ g_{\omega K}g_{\omega N} }{ m_\omega^2 }
\right) = -0.255 \mbox{ fm}
\ee
where the experimental value has been taken from \cite{Bar94}.
It can be used to fix $g_{\sigma K}$ for known $g_{\omega K}$.
Nevertheless, the presented Lagrangian is not able to get the correct
scattering length for KN-scattering in the two isospin channels.
A way out is, in addition to the $\rho$-meson,
to introduce an additional coupling of the kaon to the
scalar isovector meson $a_0$ which we will denote in the following as
$\delta$. The KN-scattering lengths for a given Isospin $I$
on the tree level are then given by
\bea
a_0^{I=1} &=& \frac{m_K}{4\pi \left( 1+ m_K/m_N \right)} \left(
\frac{ g_{\sigma K}g_{\sigma N} }{ m_\sigma^2 }
+ \frac{ g_{\delta K}g_{\delta N} }{ m_\delta^2 }
-2\frac{ g_{\omega K}g_{\omega N} }{ m_\omega^2 }
-2\frac{ g_{\rho K}g_{\rho N} }{ m_\rho^2 }
\right) \label{eq:kna0i1} \\
a_0^{I=0} &=& \frac{m_K}{4\pi \left( 1+ m_K/m_N \right)} \left(
\frac{ g_{\sigma K}g_{\sigma N} }{ m_\sigma^2 }
-3\frac{ g_{\delta K}g_{\delta N} }{ m_\delta^2 }
-2\frac{ g_{\omega K}g_{\omega N} }{ m_\omega^2 }
+6\frac{ g_{\rho K}g_{\rho N} }{ m_\rho^2 }
\right) \label{eq:kna0i0}
\quad .
\eea
The importance of the $\delta$-meson exchange contribution can be seen
if one first sets it to zero and take SU(6)-symmetry for the vector coupling
constants. Then the $a_0^{I=0}$ scattering
length depends only on the term coming from the $\sigma$-meson exchange
as $g_{\omega K}g_{\omega N} = 3 g_{\rho K}g_{\rho N}$.
This gives a large positive scattering length
whereas experiment tells us that
$a_0^{I=0} = -0.09$~fm is slightly negative \cite{Bar94}.
Taking $g_{\sigma N}=10$, $g_{\omega N} = 13$ as standard values
for the RMF model (see table~1) one gets
$a_0^{I=0}\approx 0.4$~fm without the $\delta$-meson contribution.
Including the $\delta$-meson term and using
$g_{\delta N}=5.95$ from the Bonn model \cite{Mach87}
one can fit both scatterings lengths nicely for
\be
g_{\sigma K} \approx  1.9 - 2.3 \; ,  \quad
g_{\delta K}  \approx 5.6 - 6.4
\label{eq:kncoupl}
\quad
\ee
for the various parameter sets used. A complete list of the coupling constants
of the kaon can be found in table 2.
Note that the values are surprisingly close to the one expected from the simple
quark-model ($g_{\sigma K} = g_{\sigma N}/3 \approx 3.3$,
$g_{\delta K} = g_{\delta N} = 5.95$).
The effects of the $\delta$-meson on the properties of neutron star
matter were discussed in the previous section and
seemed to be quite insigificant. Hence, despite of the rather large coupling
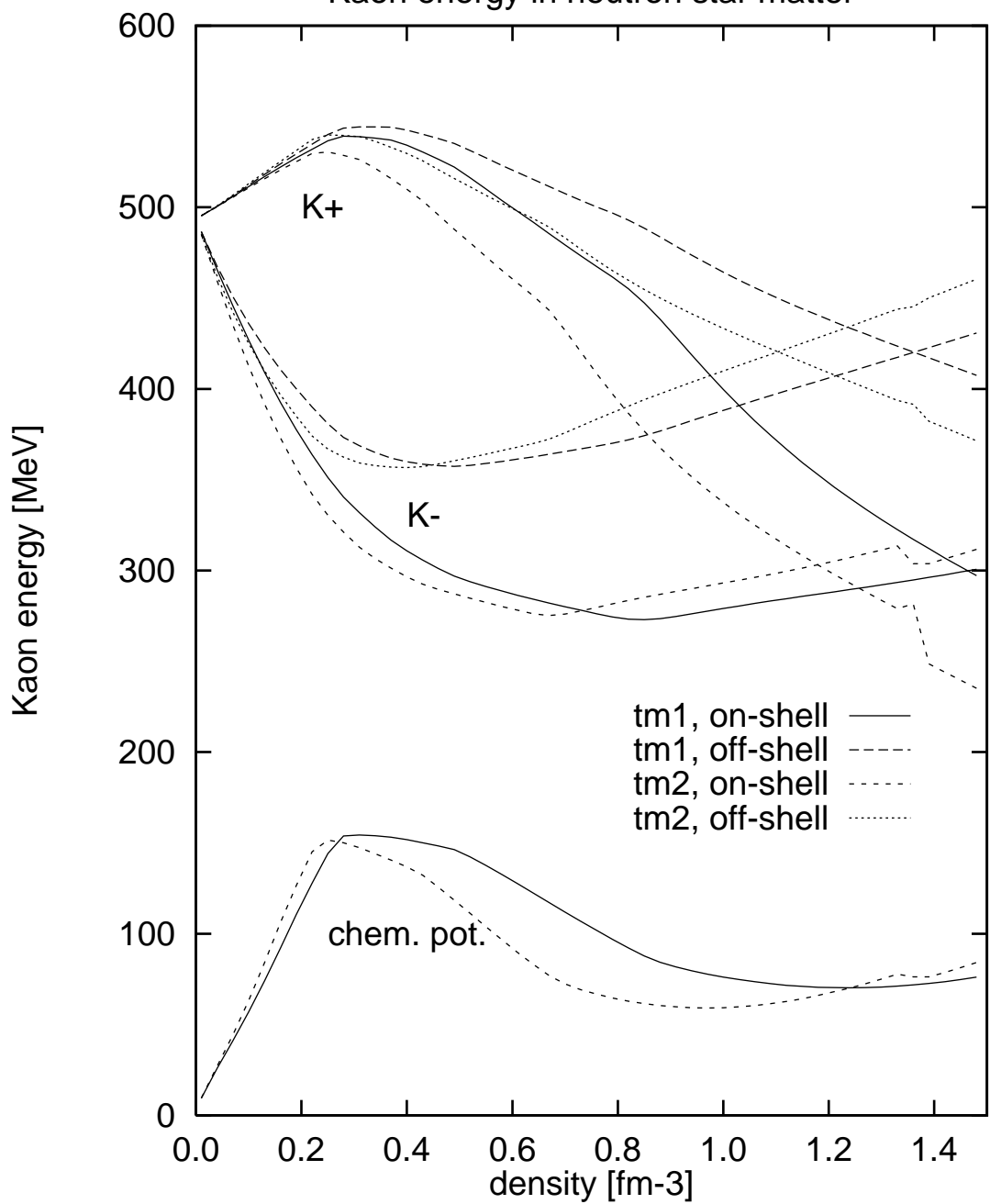
constant to the $\delta$-meson its influence on the kaon effective energy
in neutron star matter
will be rather small compared with the other isoscalar fields.

In addition, the importance of
off-shell corrections has been pointed out in detail in \cite{Delo92}.
The scalar term has to change sign when going from on-shell
$\omega_K = m_K$ to the off-shell point $\omega_K =0$.
In the one-boson exchange picture this can be incorporated by adding the
following terms
\be
{\cal L}_K'' =
D^*_\mu \bar K D^\mu K \left( \frac{f_{\sigma K}\sigma}{m_K}
+ \frac{f_{\sigma^* K}\sigma^*}{m_K} \right)
+ D^*_\mu \bar K \tau D^\mu K \frac{f_{\delta K}\vec{\delta}}{m_K}
\quad.
\ee
The on-shell constraints eqs.~(\ref{eq:kna0i1}) and (\ref{eq:kna0i0})
are still fulfilled if one replaces
$g_{\sigma K}$ by $g_{\sigma K} - f_{\sigma K}$ and
$g_{\delta K}$ by $g_{\delta K} - f_{\delta K}$, respectively, for the
values given in (\ref{eq:kncoupl}).
The off-shell behaviour of the scalar contribution
is now proportional to
$$
\propto
\left( 1-\frac{f_{\sigma K}}{g_{\sigma K}} \frac{\omega_K^2}{m_K^2}
\right) \frac{ g_{\sigma K}g_{\sigma N} }{m_\sigma^2}m_K
$$
and changes sign for $\omega = m_K/\sqrt{2}$, as it should be \cite{Delo92},
if one takes $f_{\sigma K} = 2g_{\sigma K}$.
Hence, the decrease of the effective
mass of the kaon in the nuclear medium is reduced by these new terms
\be
{m^*_K}^2 = \frac{m_K^2 + g_{\sigma K}\sigma m_K}{1+f_{\sigma K}\sigma /m_K}
\ee
and makes kaon condensation even less favourable at higher densities
(note that both the $\sigma$-field and the coupling constants are negative
here).

We have found that the fit based on the adjustment to the KN-scattering lengths
leads to an optical potential around
$U^{\bar K}_{\rm opt} = -(130\div 150)$ MeV at
normal nuclear density for the parameter sets used.
This is between the two
families of solutions found in \cite{Fried93}.

The general expression for the effective energy of the charged
kaon and the antikaon in neutron star matter is given by
\be
\omega_{K^+, K^-} = m_K \sqrt{
\frac{m_K + g_{\sigma K}\sigma + g_{\sigma^* K}\sigma^* \pm g_{\delta K}\delta}
{m_K + 2(g_{\sigma K}\sigma + g_{\sigma^* K}\sigma^* \pm g_{\delta K}\delta)} }
\;\pm\; \left( g_{\omega K} V_0 + g_{\phi K} \phi_0 + g_{\rho K} R_{0,0}
\right)
\ee
and shown in fig.~8 for the parameter sets TM1 and TM2
and model 2 including the $\delta$-meson.
The two cases (with and without the off-shell terms) are plotted.
The energy of the kaon is first increasing due to the low density theorem.
The energy of the antikaon is decreasing steadily.
After hyperons are present the situation changes dramatically.
The potential coming from the $\phi$-field cancels the contribution
coming from the $\omega$-meson. Hence, at a certain density the energies of
the kaons and antikaons are equal to the effective mass of the kaon, i.e.\
the curves for kaons and antikaons are crossing at sufficent high densities
and the energy of the kaon gets even lower than that of the antikaon!
The difference between the two parameter sets considered is not as
important as the inclusion of the off-shell terms. These terms effectively
shifts the effective mass of the kaon to higher values. Therefore the in-medium
energy of both kaons and anitkaons is shifted up to at most 100 MeV.
The minimum energy of the antikaon reads about 270 MeV for the on-shell
and 350 MeV for the off-shell case.
As the chemical potential never reaches values above 160 MeV here (or 200 MeV
without the $\delta$-meson) antikaon condensation does never occur.
We have checked the possibility of antikaon condensation for all parameter
sets for model 1 and model 2 and found that at least 100 MeV are missing
for the onset of condensation.
The most uncertain term for the effective energy of the antikaon is
the one coming from the $\sigma^*$-exchange. If one arbitrarily doubles
the coupling constant this would lead at maximum to an additional attraction
of about $70-80$ MeV and still no antikaon condensation is possible within
our approach.

\section{Conclusion and Outlook}

We have studied the equation of state of neutron star matter within
the relativistic mean field model including hyperons. We have used
a representative set of parameters fitted to the properties of finite nuclei
and included the rather strong hyperon-hyperon interaction induced by
additional
(hidden) strange meson fields.
The coupling constants of the hyperons have been fixed by SU(6)-symmetry
relations and hypernuclear properties.
Standard parametrizations with different forms of the scalar selfinteraction
yield negative values for the nucleon effective mass solely due to
the abundant presence of hyperons at high densities.
The vector selfinteraction terms and the additional meson
exchanges together eliminate this behaviour.
According to our calculations,
the baryon effective masses become small
in neutron star matter at very high densities ($\rho > 6\rho_0$)
and the abundances of the different species are determined
by their spin-degeneracy factors (i.e.\ isospin-saturated matter).
In all cases the hyperons get into the game at quite low densities, around
$2-3\rho_0$, and the composition is quite similar for all parameter sets
studied up to $4\rho_0$.
The effects coming from the inclusion of a scalar isovector meson
$a_0(980)$ have been discussed.
The equation of state does not change considerably
but all the baryons get now different effective masses.
The protons and electrons are getting a little bit suppressed at lower
densities and the elctrochemical potential decreases
while the $\Sigma^-$'s occur at slightly lower densitites compared
to the model without the $\delta$-meson contribution.

The possibility of antikaon condensation has been also studied in the framework
of the relativistic mean-field model.
Fixing the vector coupling constants by SU(3)-constraints, the
in-medium energy of the antikaon was calculated for different
choices of the optical potential at normal nuclear density.
The existence of hyperons, especially the repulsive contribution from
$\phi$-exchange and the decrease of the electrochemical potential,
makes a antikaon condensed phase less favourable even at very high densities.
We get different answers for different parameter sets whether or not
antikaons are present in the dense interior of neutron stars.
Nevertheless, we see no onset of antikaon condensation
below $\rho < 6 \rho_0$ for all cases studied.
Note that the maximum density of neutron star will set another limit for
the kaon condensed phase which has not been studied here.

In a second approach we fix the coupling constants to the
available KN-scattering data and take care of the off-shell behaviour
of the kaon energy.
Hence, this approach fulfills the low-density theorem for kaons.
The effective energy of the kaons and antikaons are dictated by the behaviour
of the effective mass of the kaon as the contributions from the vector fields
cancel each other. For very high densities ($\rho >7\rho_0$) it is possible
that the effective energy of the kaon is even lower than that of the antikaon
as the contribution form the (hidden) strange meson field exceds the one
coming from the $\omega$-meson field. The effective energy
of the antikaon turns upward again at this density.
Hence, antikaon condensation can never occur in any of our
parametrizations as the electrochemical potential is always lower than
the minimum energy of the antikaon.

The extrapolation to hyperon-rich matter is associated with large
uncertainties due to the unknown hyperon-kaon interaction
(e.g.\ $\bar K\Xi$-interaction).
The effective energy of the kaon is very much influenced by large
cancellation effects coming from the vector fields.
A slight increase of the coupling
constant might therefore change the antikaon energy by 100 MeV
which are at least missing for the onset of kaon condensation.
Hence, there might be still some
niches for antikaon condensation. Nevertheless, the overall trend
is quite clear: the presence of hyperons makes the onset of antikaon
condensation quite unlikely to happen.

The vector selfinteraction terms as well as the hyperon-hyperon interactions
might influence also the maximum mass,
the rotational frequency and cooling properties
of neutron stars. These questions will be studied in a forthcoming work.

\section*{Acknowledgements}

We would like to thank J. Bondorf, N.K. Glendenning, V. Thorsson, F. Weber,
W. Weise and Th. Waas for useful discussions and remarks.
This work is supported in part by the International Science Foundation (Soros)
grant N8Z000 and EU-INTAS grant 94-3405.
J. Schaffner is supported by the EU-program 'Human Capital and Mobility'
(contract ERBCHBGCT930407).
The authors also thank the Niels Bohr
Institute for kind hospitality and financial support.

\begin{table}

\caption{The coupling constants of the parameter sets used.
The vector coupling constants for the hyperons are be taken from
SU(6)-relations. The nucleons do not couple to the $\sigma^*$- and the
$\phi$-meson ($g_{\sigma^* N}=g_{\phi N} = 0$). The $\Lambda$ does not couple
to the isovector fields ($g_{\rho \Lambda}=0$).
The coupling constants for the $\Sigma$'s are the same as
for the $\Lambda$ except for the isovector coupling constant which
is $g_{\rho\Sigma}=2g_{\rho N}$.
The parameters for the scalar and vector selfinteraction
terms are not given, they can be found in the corresponding references.}
\begin{center}
\begin{tabular}{ccccccc}
Set & nl-z  &    nl-sh &    pl-z  &    pl-40 &    tm1   &   tm2   \cr
\hline
Ref. & \cite{Rufa88} & \cite{Shar93} & \cite{Rei88} & \cite{Rei88} &
\cite{Toki94} & \cite{Toki94} \cr
\hline
$g_{\sigma N}$ &          10.0553 &  10.4440 &  10.4262 &  10.0514 &  10.0289 &
11.4694 \cr
$g_{\omega N}$ &          12.9086 &  12.9450 &  13.3415 &  12.8861 &  12.6139 &
14.6377 \cr
$g_{\rho N}$ &             4.8494 &   4.3830 &   4.5592 &   4.8101 &   4.6322 &
4.6783 \cr
\hline
$g_{\sigma \Lambda}$ &     6.23 &     6.47 &     6.41 &     6.20 &     6.21 &
7.15 \cr
$g_{\omega \Lambda}$ &     8.61 &     8.63 &     8.89 &     8.59 &     8.41 &
9.76 \cr
$g_{\sigma^* \Lambda}$ &   6.77 &     6.85 &     6.93 &     6.78 &     6.67 &
7.65 \cr
$g_{\phi \Lambda}$ &      -6.09 &    -6.10 &    -6.29 &    -6.07 &    -5.95 &
-6.90 \cr
\hline
$g_{\sigma \Xi}$ &         3.45 &     3.59 &     3.52 &     3.43 &     3.49 &
3.94 \cr
$g_{\omega \Xi}$ &         4.30 &     4.31 &     4.45 &     4.30 &     4.20 &
4.88 \cr
$g_{\rho \Xi}$ &           4.85 &     4.38 &     4.56 &     4.81 &     4.63 &
4.68 \cr
$g_{\sigma^* \Xi}$ &      12.59 &    12.66 &    12.95 &    12.57 &    12.35 &
14.18 \cr
$g_{\phi \Xi}$ &         -12.17 &   -12.20 &   -12.58 &   -12.15 &   -11.89 &
-13.80
\end{tabular}
\end{center}
\end{table}

\begin{table}
\caption{The coupling constants for the kaons.
The coupling constants to the $\sigma$- and $\delta$-meson are fixed
by the s-wave KN-scattering lenghts. The vector coupling constants
are chosen from SU(3)-relations. The coupling constant to the $\sigma^*$-meson
is taken from $f_0$-decay \protect\cite{Arm91}.}
\begin{center}
\begin{tabular}{ccccccc}
Set & nl-z  &    nl-sh &    pl-z  &    pl-40 &    tm1   &   tm2   \cr
\hline
Ref. & \cite{Rufa88} & \cite{Shar93} & \cite{Rei88} & \cite{Rei88} &
\cite{Toki94} & \cite{Toki94} \cr
\hline
$g_{\sigma K}$   &  1.85 &  2.05 &   2.20 &   2.27 &   1.93 &   2.27 \cr
$g_{\omega K}$   &  3.02 &  3.02 &   3.02 &   3.02 &   3.02 &   3.02 \cr
$g_{\rho K}$     &  3.02 &  3.02 &   3.02 &   3.02 &   3.02 &   3.02 \cr
$g_{\sigma^* K}$ &  2.65 &  2.65 &   2.65 &   2.65 &   2.65 &   2.65 \cr
$g_{\phi K}$     &  4.27 &  4.27 &   4.27 &   4.27 &   4.27 &   4.27 \cr
$g_{\delta K}$   &  6.37 &  5.59 &   5.89 &   6.31 &   5.87 &   5.94
\end{tabular}
\end{center}
\end{table}

\figure{FIG. 1.
The composition of neutron star matter with hyperons
in model 2 using parameter set PL-Z.
The jump in the curves is due to the negative effective mass of the nucleon.
}

\figure{FIG. 2.
The composition of neutron star matter with hyperons
in model 2 using parameter set TM1 with vector selfinteraction terms.
The matter approaches isospin-saturation at high densities.
}

\figure{FIG. 3.
The effective masses of the baryons versus the density for model 2
using set TM1.
All effective masses remain positive for the density range considered.
}

\figure{FIG. 4.
The field potentials and the electrochemical potential versus the density
for model 2 using set TM1.
Quite high values are reached indicating large cancelation effects at high
densities.
}

\figure{FIG. 5.
The equation of state in model 2 for various parameter sets.
The causal limit $p=\epsilon$ is also shown.
}

\figure{FIG. 6.
The effective masses of the baryons versus the density for model 2
using set TM1 with $\delta$-mesons.
All baryons get a different effective mass due to the
additional meson.
}

\figure{FIG. 7.
The effective energy minus the chemical potential of the $K^-$
over the density for model 2 using various parameter sets.
The upper curves are calculated for an optical potential of
$U^{\bar K}_{\rm opt}= -120$~MeV, the lower ones for
$U^{\bar K}_{\rm opt}= -200$~MeV.
}

\figure{FIG. 8.
The effective energy of the kaon and the antikaon
for model 2 with $\delta$-mesons using set TM1.
The electrochemical potential is also plotted.
Antikaon condensation does not occur over the whole density region
considered.
}

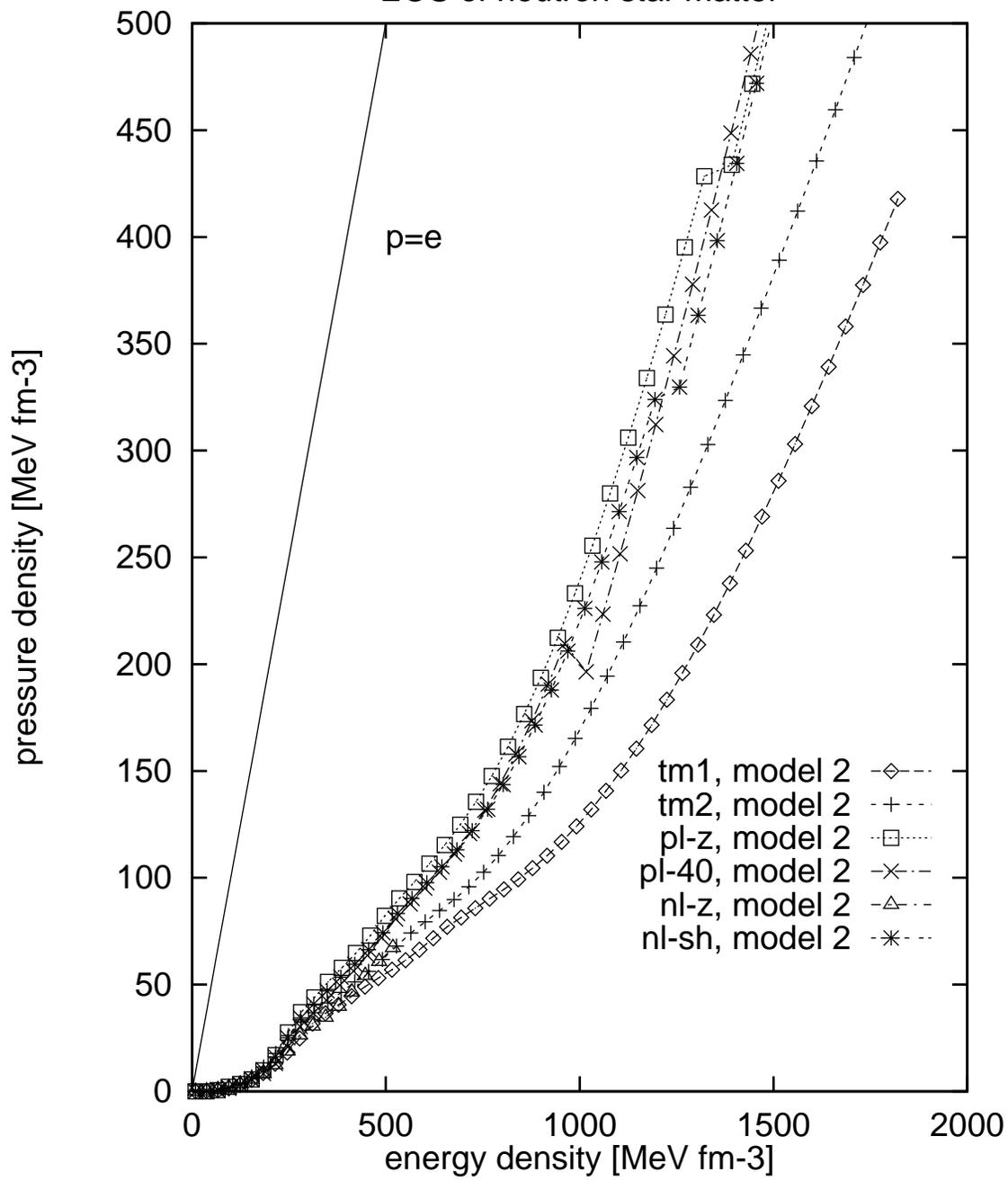
\end{document}